\documentclass{article}
\usepackage{amsmath,graphicx,booktabs,color,cite}
\usepackage{spconf}
\usepackage{amssymb}
\usepackage{amsbsy}
\usepackage{bm}
\usepackage{url}
\usepackage{multirow}
\usepackage{subcaption}
\usepackage{hyperref}
\title{SiFiSinger: A High-Fidelity End-to-End Singing Voice Synthesizer based on Source-filter Model
}
\name{Jianwei Cui$^{1,2}$, Yu Gu$^2$, Chao Weng$^2$, Jie Zhang$^1$, Liping Chen$^1$, Lirong Dai$^1$\thanks{This work was done at Tencent AI Lab as an internship by Jianwei Cui.}}
\address{$^1$NERC-SLIP, University of Science and Technology of China (USTC), Hefei, China \\
         $^2$Tencent AI Lab\\}

\begin{document}
\ninept
\maketitle
\begin{abstract}
This paper presents an advanced end-to-end singing voice synthesis (SVS) system based on the source-filter mechanism that directly translates lyrical and melodic cues into expressive and high-fidelity human-like singing.
Similarly to VISinger 2, the proposed system also utilizes training paradigms evolved from VITS and incorporates elements like the fundamental pitch (F0) predictor and waveform generation decoder.
To address the issue that the coupling of mel-spectrogram features with F0 information may introduce errors during F0 prediction, we consider two strategies. Firstly,
we leverage mel-cepstrum (mcep) features to decouple the intertwined mel-spectrogram and F0 characteristics.  Secondly, inspired by the neural source-filter models, we introduce source  excitation signals as the representation of F0 in the SVS system, aiming to capture pitch nuances more accurately. 
Meanwhile, differentiable mcep and F0 losses are employed as the waveform decoder supervision to fortify the prediction accuracy of speech envelope and pitch in the generated speech. Experiments on the Opencpop dataset  demonstrate efficacy of the proposed model in synthesis quality and intonation accuracy.
Synthesized audio samples are available at: \url{https://sounddemos.github.io/sifisinger}.
\end{abstract}
\vspace{-0.1cm}
\section{Introduction}\label{sec:intro}
% \vspace{-0.1cm}
Singing voice synthesis (SVS)
aims to fabricate realistic singing voices given song lyrics and musical scores which contain different kinds of musical features such as note and tempo information.
Recently, SVS techniques have developed significantly, mainly owning to breakthroughs in deep learning \cite{gu2021bytesing,ren2020deepsinger,zhang2022susing,liu2022diffsinger,blaauw2020sequence,hono2021sinsy}.
Contemporary SVS systems usually adopt a two-stage approach, where an acoustic model first analyzes the musical score to extract lyrical and musical information and predicts acoustic features such as the mel-spectrogram (mel), followed by the usage of a separate neural vocoder that translates these features into audible waveforms \cite{gu2021bytesing, liu2022diffsinger}.

Different from such a two-stage scheme, VISinger \cite{zhang2022visinger} and VISinger 2 \cite{zhang23e_interspeech} are fully end-to-end approaches for SVS, which can generate singing waveforms directly given input music scores. On the basis of the main architecture of VITS \cite{kim2021conditional}, VISinger 2 also employs variational autoencoder (VAE) based posterior encoder, prior encoder and adversarial decoder. Equipped with the length regulator and  F0 predictor, VISinger 2 models the frame-level mean and variance of both the prior and posterior distributions related to acoustic features during encoding. Besides, VISinger 2 uses a differentiable DSP (DDSP) synthesizer \cite{engel2019ddsp}, with a harmonic synthesizer and a noise synthesizer to  model the harmonic and aperiodic audio components from the latent representation, respectively. The periodic and aperiodic signals obtained from the DDSP synthesizer are concatenated to serve as the conditional input for the adversarial decoder. Their summation is then used to produce a waveform for calculating the loss function. This approach helps alleviate the text-to-phase issue and enhances the modeling capability of the system.
%Such pioneering methods facilitate superior control over the synthesized sound's attributes.

Nevertheless, the VISinger 2 still suffers from some issues in the context of SVS. For example, pitch accuracy is more crucial for SVS than text-to-speech (TTS), since F0s are  directly relevant to music notes in SVS rather than linguistic lyrics. Mel-spectrograms which are commonly-used as acoustic features for both TTS and SVS tangle F0 with spectral envelope, and VISinger 2 thus utilizes predicted F0 information as a prior condition to predict the mel-spectrograms. An inherent coupling between the mel-spectrograms and F0 information can lead to error propagation. Biases in F0 modeling may adversely influence the prediction accuracy of mel-spectrograms. 
Although the DDSP synthesizer 
 in VISinger 2 can ease the text-to-phase modeling, it lacks consistent and direct utilization of F0 information in prior hidden vector prediction and  subsequent audio waveform generation.
 Such issues culminate in synthesis quality descent and pitch inaccuracy in the final audio output.

\begin{figure*}[t]
    \vspace{-1cm}
    \centering
    \includegraphics[width=\textwidth, keepaspectratio]{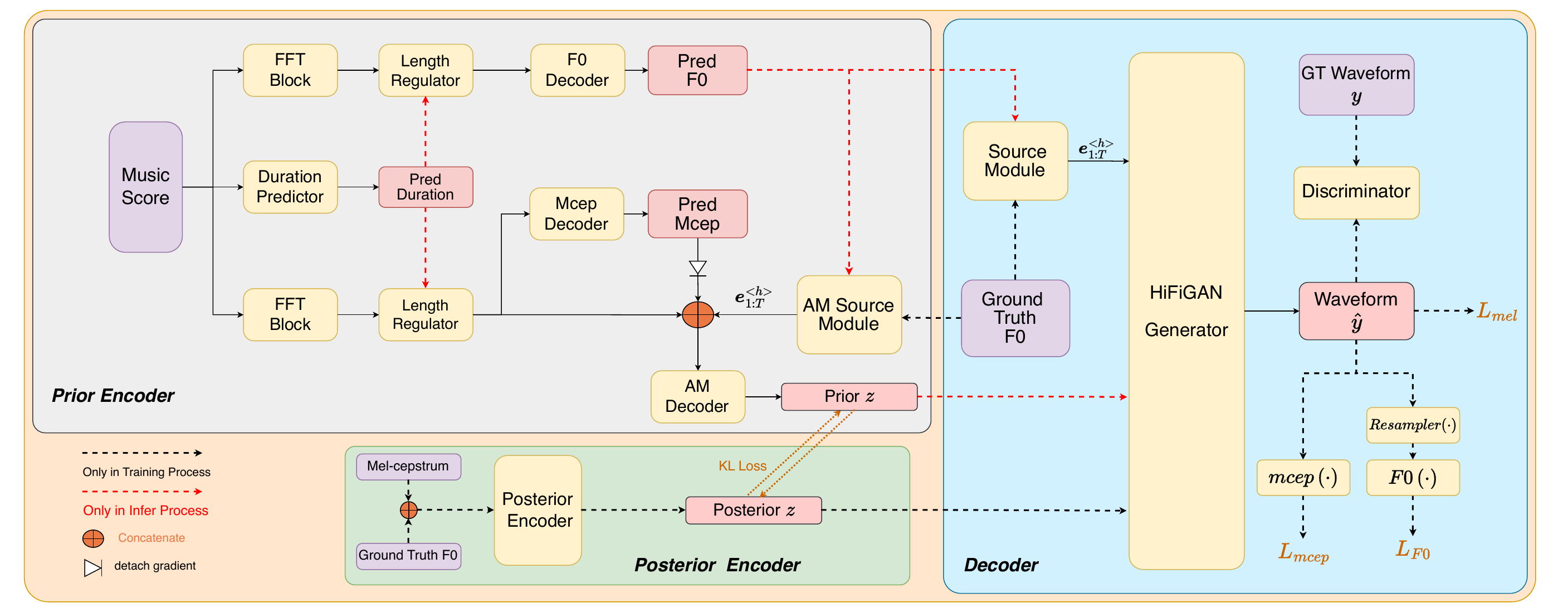}
    \caption{The overview of SiFiSinger, where the F0 decoder, Mcep decoder and AM (Acoustic Model) decoder are FFT blocks, the black dashed lines indicate operations that only occur during training, and the red dashed lines indicate operations that are considered only for inference.}
    \label{fig1:archtecture}
\end{figure*}

In this work, we therefore propose a SiFiSinger SVS system based on the source-filter mechanism of speech production. The source-filter model \cite{harrington1999acoustic} is a foundational model in voice science that describes the human voice generation mechanism. The ``source" usually pertains to the vibrations of the vocal cords producing the foundational sound or waveforms. The ``filter" can be envisioned as the process through which the voice produced by the source travels via the vocal tract. Motivated by 
neural source-filter model \cite{wang2019neural,10095298}, the F0s are processed using the source module in the SiFiSinger prior encoder, generating multiple harmonics controlled by F0. Additionally, the F0 excitation processed by the source module is also utilized as a pitch embedding in the HiFi-GAN \cite{kong2020hifi} generator within the decoder, ensuring superior pitch control during sound waveform generation. 
Furthermore, we utilize the mcep features \cite{1163420} to replace mel features, which capture spectral envelope information, decoupled from F0 and phase information. This can be perceived as the ``filter" component in the source-filter theory. By concatenating the mcep feature with the source excitation signal generated by F0, the modeling process of the acoustic features in the prior and posterior encoders can be viewed as a variant of neural source-filter model. 
In addition, we use a differentiable method to re-extract mceps and F0s on the synthesized audio of the  generator, and calculate the loss with the ground truth. Benefiting from this, the gradient  can be back-propagated to the generator. 
This creates a more direct and effective separated supervision with ``source" F0 and ``filter" mcep to the whole model.
Experiments demonstrate that SiFiSinger exhibits better synthesized audio quality and pitch accuracy than VISinger 2.
\vspace{-0.2cm}
\section{methodology}
\vspace{-0.2cm}
\label{sec:method}
The {overall architecture of the proposed SiFiSinger is illustrated in Fig.~\ref{fig1:archtecture}, adopting a structure reminiscent of VISinger 2 and VITS with a conditional VAE framework. This structure encompasses a prior encoder, a posterior encoder and a waveform decoder. In this section, we will introduce each module in detail}.

% This structure encompasses a prior encoder, a posterior encoder and a waveform decoder. And source modules which can generate the source excitation signal controlled by F0 are utilized throughout these modules to better model the musical pitch information.
% The subsequent sections will delve into a detailed exposition of each constituent part.
% This structure encompasses a prior encoder, a posterior encoder, and a decoder. Throughout the acoustic modeling phase, the source excitation signal controlled by F0 is consistently utilized, with varied applications. The subsequent sections will delve into a detailed exposition of each constituent part.
\vspace{-0.2cm}
\subsection{Source module}
\label{source module}
The source module \cite{wang2019neural} is designed to produce a sinusoidal excitation, say \( \boldsymbol{e}_{1:T} = \{\boldsymbol{e}_1,...,\boldsymbol{e}_T\} \), using the F0 sequence \( \boldsymbol{f}_{1:T} \), {where $e_t\in\mathbb{R}$, $ t \in \{1,...,T\}$, with \( t \) representing} the \( t \)-th time step. The generation of the sinusoidal excitation \( \boldsymbol{e}_{1: T}^{<0>} \) can be {formulated as}
\begin{equation}
    \boldsymbol{e}_t^{<0>}= \begin{cases}\alpha\textit{sin}\left(\sum_{k=1}^t 2 \pi \frac{f_k}{N_s}+\phi\right)+n_t, & \text { if } f_t>0 \\ \frac{1}{3 \sigma} n_t, & \text { if } f_t=0\end{cases}\label{eq:source model}
\end{equation}
where $n_t \sim \mathcal{N}\left(0, \sigma^2\right)$, \(\phi\) is a random initial phase, and \(N_s\) is the waveform sampling rate. The hyper-parameter \(\alpha\)
 adjusts the amplitude of source waveforms, while \(\sigma\) is the standard deviation of the Gaussian noise.
The source module also generates harmonic overtones, {where  the \( h \)-th harmonic overtone corresponds to the \( \left(h+1\right) \)-th harmonic frequency. As such, the sinusoidal excitation \( \boldsymbol{e}_{1: T}^{<h>} \) is obtained using \( \left(h+1\right)f_t \) in  (\ref{eq:source model}). In the final step, the source module employs a trainable feed forward (FF) layer to merge \( \boldsymbol{e}_{1: T}^{<0>} \) and \( \boldsymbol{e}_{1: T}^{<h>} \) together. By applying the F0-controlled harmonic generation mechanism, the module can ensure that the synthesized voice aligns closely with the intended pitch and enriches the overall quality and naturalness of the resulting} singing voice.
% $\boldsymbol{f}_{1:T}$,$\boldsymbol{e}_{1:T} = \{e_1,...,e_T\}$,$e_t\in\mathbb{R}$,
% $\forall{t} \in \{1,...,T\}$
% $n_t \sim \mathcal{N}\left(0, \sigma^2\right)$
% $\boldsymbol{e}_{1: T}=\boldsymbol{e}_{1: T}^{<0>}$
\vspace{-0.2cm}
\subsection{Prior encoder}
The structure of the {prior encoder is largely congruent with that of VISinger 2, adopting a feed-forward Transformer (FFT) block and length regulator in FastSpeech \cite{ren2019fastspeech}. The prior encoder contains a duration predictor, F0 and mcep acoustic decoders, where both} take the music score as inputs. During the training phase, based on the ground truth duration, mcep features and F0s are generated by the acoustic decoders  following the loss function \( L_{am} \), given by
\begin{equation}
L_{am}\! = \!\lambda_{1}\textit{MSE}\left(LF0, LF0_{pred}\right) + \lambda_{2}\|Mcep - Mcep_{pred}\|_1,
\end{equation}
where \( LF0_{pred} \) and \( Mcep_{pred} \) are the predicted log-F0 and mcep, $\lambda_{1}$,$\lambda_{2}$ are the corresponding coefficients, respectively, and $\textit{MSE}(\cdot)$ represents mean squared error loss. Here, the mcep feature is employed, which primarily captures the audio envelope information without being entangled with F0.
The F0 is processed through the source module in Section \ref{source module} to provide a series of sinusoidal harmonics controlled by F0. This excitation signal, which represents the rapidly oscillating periodic harmonic components in the audio, is concatenated with the mcep feature. The spectral envolope feature and pitch information are modeled separately to reflect the characteristics of the vocal tract and vocal cords, respectively, which aim to imitate the pronunciation mechanism of human singing. The duration predictor forecasts the duration of both phonemes and notes and computes the duration loss \( L_{dur} \). During the inference phase, the length regulator relies on the output of the duration predictor as a length reference. 
In the end, {based on the output of the music score encoder, the mcep feature and the excitation signal from the AM (Acoustic Model) source module, the mean and variance of the prior distribution at the frame level is predicted by the AM decoder. The prior hidden vector \( z \) can then be sampled as well}.

\begin{figure*}[t]
    % \vspace{-0.5cm}
    \centering % 使图像居中
    \includegraphics[width=\textwidth]{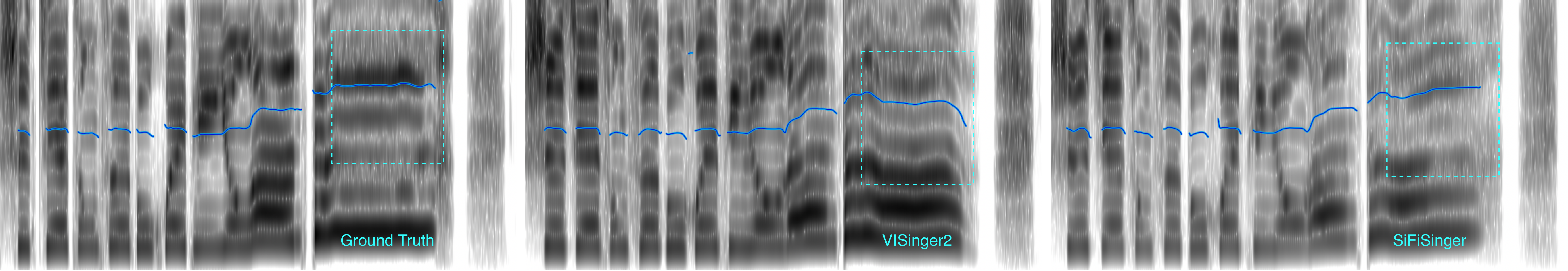} % 设置图像宽度为单列宽度
    \caption{Visualization of spectrums and pitch contours obtained by ground truth (left), VISinger 2 (middle) and SiFiSinger (right).} % 添加图像标题
    \label{fig:pitch} % 为图像添加标签，以便在文档其他地方引用
\end{figure*}
\vspace{-0.2cm}
\subsection{Posterior encoder}
{Based on the backbone of VISinger 2, 
the posterior encoder consists of} \( N \) 1-D convolution layers and LayerNorm layers. Different {from} VISingers,  we directly use the mcep feature and F0 as the posterior encoder inputs. Similar{ly to the} prior encoder, it predicts the mean and variance of the posterior distribution given the  frame-level acoustic features. Subsequently, a re-sampling procedure is applied to obtain the posterior latent vector \( z \). During the training phase, {this posterior \( z \) and the prior \( z \) are} constrained via the KL divergence \cite{kullback1951information} loss \( L_{kl} \).
\vspace{-0.2cm}
\subsection{Decoder}
The decoder consists of a HiFi-GAN generator, {which utilizes} the latent distribution \( z \) as input and produces the waveform \( \hat{y} \). We extract the mel-spectrogram from the synthesized waveform \( \hat{y} \) to calculate the mel-spectrogram loss \( L_{mel} \).Within the decoder, we continue to employ the excitation signal generated by the source module to provide strong pitch information during the generation process. Unlike the prior encoder, the frame-level F0 is first upsampled  to sample-level and then passed through the source module to produce the excitation signal. This is then progressively downsampled and combined with the progressively upsampled \( z \) within the generator, offering multi-scale pitch information during the waveform generation process. We adopt an adversarial learning approach \cite{goodfellow2014generative}, utilizing a discriminator \( D \) to distinguish between the waveform \( \hat{y} \) produced by the generator \( G \) and the ground truth waveform \( y \). In the adversarial training, both the least-squares loss and feature-matching loss are considered as
\begin{gather}
L_{adv}(D)  = \mathbb{E}_{(y, z)}\left[(D(y)-1)^2 + (D(G(z)))^2\right],\label{eq:LadvD} \\
L_{adv}(G)  = \mathbb{E}_z\left[(D(G(z))-1)^2\right],\label{eq:LadvG} \\
L_{fm}(G)  = \mathbb{E}_{(y, z)}\left[\sum_{l=1}^T \frac{1}{N_l} \left\|D^l(y) - D^l(G(z))\right\|_1\right],\label{eq:Lfm}
\end{gather}
where \( T \) represents the number of layers in the discriminator, \( l \) stands for the \( l \)-th layer of the discriminator, and \( N_l \) denotes the number of features in the \( l \)-th layer, the discriminator's architecture and configurations are the same as those in VISinger 2. The generator loss \( L_G \) is thus defined as follows:
\begin{gather}
L_G = L_{adv}(G) + \lambda_{mel}L_{mel} + \lambda_{fm}L_{fm}(G)\label{eq:6}.
\end{gather}
\subsection{Differentiable reconstruction loss}
{In addition, we use} a trained  {CREPE}  \cite{kim2018CREPE} and \texttt{diffsptk}\footnote{\url{https://github.com/sp-nitech/diffsptk}}, \ a differentiable version of the speech signal processing toolkit (SPTK) \cite{sp-nitech2023sptk} to re-extract the F0 and mcep from the waveform \( \hat{y} \) generated by the generator in a differentiable manner. {CREPE}  is a F0 estimation method that utilizes a simple convolutional neural network model, trained to predict pitch directly from raw audio waveforms, which eliminates the need for complex feature or signal pre-processing and can deal with raw audio directly. The {CREPE} model outputs a probability distribution of the audio pitch. In the original implementation of the CREPE model, decoding methods similar to argmax or Viterbi \cite{viterbi1967error} are used to obtain the predicted F0 value based on the probability distribution. However, these methods introduce some non-differentiable operations that block the backpropagation of gradients. We reconstruct the {non-differentiable operations} in the original code of {CREPE} to make sure the {back-propagation of gradients} to the input waveform. Specifically, we perform a weighted sum on the predicted pitch probability distribution over the corresponding frequency scale to obtain the final predicted F0 value. Since the {CREPE} model is trained {with a sampling rate of 16kHz,} we first resample both the generator-produced waveform \( \hat{y} \) and the ground truth \( y \) to 16kHz, yielding \( \hat{y_{\text{rs}}} \) and \( y_{\text{rs}} \), respectively. We then apply the aforementioned method to re-extract the F0 from both \( \hat{y_{\text{rs}}} \) and \( y_{\text{rs}} \) and calculate the loss accordingly. For the mcep feature, \texttt{diffsptk} implements a series of differentiable operations such as framing, windowing, and short-time Fourier transform, etc. This allows us to differentiably extract mcep features from the synthesized waveform $\hat{y}$ and compute the loss with the ground truth mcep.
 The associated reconstruction losses are computed as:
\begin{gather}
y_{rs}=Resampler(y), \hat{y}_{rs}=Resampler(\hat{y}), \\
L_{F0} = \lambda_{f0} * \textit{MSE}(F0(y_{rs}), F0(\hat{y}_{rs})), \label{eq:Lf0} \\
L_{mcep} = \lambda_{mcep} * \|mcep(y) - mcep(\hat{y})\|_1, \label{eq:Lmcep}  
\end{gather}
where  \( F0(\cdot) \) and \( mcep(\cdot) \) represent the {functions of re-extracting the  F0 and mcep feature using {CREPE} and \texttt{diffsptk}, and $\lambda_{f0}$ and $\lambda_{mcep}$ are the corresponding coefficients, respectively}. \(Resampler(\cdot)\) represents the {resampling function.} Benefiting from these differentiable operations, the gradients of these two reconstruction losses can be back-propagated from the HiFi-GAN generator to other modules in SiFiSinger. {Note that} the parameters of {CREPE} are fixed for the training procedures. The total loss for the entire training procedure can be formulated as:
\begin{gather}
L = L_{G} + L_{kl} + L_{am} + L_{dur} + L_{mcep} + L_{F0}, \\
L(D) = L_{adv}(D),
\end{gather}
where \( L(D) \) denotes the discriminator loss in (\ref{eq:LadvD}), \( L_G \) and \( L_{fm}(G) \) are given by (\ref{eq:6}) and (\ref{eq:Lfm}), respectively. During the training phase, \( L \) and \( L(D) \) are alternately optimized.
\vspace{-0.2cm}
\section{EXPERIMENTS}
\label{sec:exp}
\subsection{Data preparation}
We conduct {the training and evaluation of the proposed} SiFiSinger on the Opencpop dataset \cite{wang22b_interspeech}, {which  is a publicly available} high-quality Mandarin singing corpus. It comprises 100 unique Mandarin songs performed by a professional female singer. All audio files in this dataset were recorded {in the} studio quality at a sampling rate of 44.1 kHz. The {dataset consists of 3,756 utterances and approximately 5.2 hours in total. Pre-split training and testing sets are} used in our experiments. \texttt{diffsptk} is utilized to extract 80-dimensional mcep features from the original audio files and mean-variance normalization {is then performed. F0 is} extracted using the Harvest  algorithm \cite{morise2017harvest}. The Opencpop dataset includes text annotations for utterances, notes and phonemes, as well as pitch types and boundaries. After  pre-processing, these text annotations are used as inputs.

\subsection{Experimental setup}
To  evaluate the efficacy of {the proposed method, we choose three systems} for comparison:
%d four different systems to assess their performance and conducted an ablation study. The systems are as follows:
\begin{itemize}

  \item \textit{\textbf{VISinger 2}}: We {retain the original configuration, which serves as the baseline in} experiments.
%  \item \textit{\textbf{SiFiSinger}}: The proposed system, as described in Section 2.
  \item \textit{\textbf{SiFiSinger-ds}}: {a variant  of the proposed SiFiSinger by omitting the differentiable reconstruction loss  in (\ref{eq:Lmcep}) and (\ref{eq:Lf0})}.
  \item \textit{\textbf{SiFiSinger-as}}: {a variant  of the proposed SiFiSinger by excluding the source module in the prior encoder, i.e., the AM source module in Fig.~\ref{fig1:archtecture}.}
\end{itemize}
 The hyper-parameters for SiFiSinger {are mostly kept} consistent with those used in VISinger 2 \cite{visinger_github}.  The hidden size and filter channels {are set to 192 and 768,} respectively. The posterior encoder is composed of 8 layers of 1-D convolution and normalization layers. The HiFi-GAN generator employed upsample rates of \([8,8,4,2]\) and upsample kernel sizes of \([16,16,8,4]\), with a hidden size of 256.
All models {are trained for 500k steps, where the batch size is set to 16}. Training is completed  using the AdamW  optimizer \cite{loshchilov2018decoupled} with hyper-parameters \( \beta_1 = 0.8 \) and \( \beta_2 = 0.99 \).
% \begin{figure}[h]
%     \centering
%     \begin{subfigure}[b]{\columnwidth}
%         \centering
%         \includegraphics[width=\columnwidth]{gt_pitch_2.jpg}
%         \caption{a}
%         \label{fig:subfig_a}
%     \end{subfigure}
    
%     % 添加一些垂直空间（可选）
%     % \vspace{1em}
    
%     \begin{subfigure}[b]{\columnwidth}
%         \centering
%         \includegraphics[width=\columnwidth]{baseline_pitch_2.jpg}
%         \caption{b}
%         \label{fig:subfig_b}
%     \end{subfigure}
    
%     % % 添加一些垂直空间（可选）
%     % \vspace{1em}
%     \begin{subfigure}[b]{\columnwidth}
%         \centering
%         \includegraphics[width=\columnwidth]{sifisinger_pitch_2.jpg}
%         \caption{c}
%         \label{fig:subfig_c}
%     \end{subfigure}
%     \caption{Visualization of spectral pitches}
%     \label{fig:main_figure}
% \end{figure}
\vspace{-0.1cm}
\subsection{Objective evaluation}
% \vspace{-0.1cm}
%Objective evaluations were conducted on the aforementioned systems for a comprehensive understanding of their performances. 
{In order to measure the performance, we compute} various metrics, including F0 root mean square error (F0 RMSE), mel-spectrum RMSE (Mel RMSE), F0 correlation coefficient (F0 Corr), and the error rate of voiced/unvoiced frames (V/UV). Note that for the calculation of distortions, the ground truth duration is used. The results are presented in Table~\ref{table:obj}.
\begin{table}[t]
    \centering
    \begin{tabular}{lccccc}
        \toprule
        \multirow{2}{*}{Model} & \multicolumn{1}{c}{F0} & \multicolumn{1}{c}{Mel} & \multicolumn{1}{c}{F0} & \multicolumn{1}{c}{\multirow{2}{*}{V/UV\( \downarrow \)}} \\ 
                                & RMSE\( \downarrow \)  & RMSE\( \downarrow \)  & Corr\( \uparrow \)  & \\
        \midrule
        \textbf{\textit{VISinger 2}}           & 44.17 & 0.37  & 0.756 & 5.87\%  \\
        \textbf{\textit{SiFiSinger-as}}  & 45.70 & 0.36  & 0.727 & \textbf{4.75\%} \\
        \textbf{\textit{SiFiSinger-ds}}         & 43.10 & 0.35  & 0.742 & 5.71\%  \\
        \textbf{\textit{SiFiSinger}}          & \textbf{42.93} & \textbf{0.35}  & \textbf{0.761} & 5.89\% \\
        \bottomrule
    \end{tabular}
    \caption{The objective evaluation results of different systems.}
    \label{table:obj}
\end{table}
{It is clear that the proposed \textbf{\textit{SiFiSinger}} system outperforms  \textbf{\textit{VISinger 2}} and both variants} in terms of F0 RMSE. The incorporation of the two source modules enables the model to utilize F0 information more accurately and sufficiently during the synthesis process. Moreover, the addition of the differentiable \(L_{F0}\) loss function in \textbf{\textit{SiFiSinger}} establishes a more direct form of supervision, leading to improved accuracy in F0.
Furthermore, \textbf{\textit{SiFiSinger}} also {achieves a} lower spectral distortion compared to the baseline, {i.e., a more accurate prediction of the spectral information}.
We also visualize the spectrum and  pitch contour of a {audio example in Fig~\ref{fig:pitch}, the blue solid line depicts the pitch of the audio in a logarithmic frequency scale. From the figure, we can clearly see that the true audio ends in a rising tone. The \textbf{\textit{VISinger 2}} produces a completely inaccurate descending tone, but} \textbf{\textit{SiFiSinger}} can match the pitch contour of the ground truth more closely, {showing the superiority in the} pitch accuracy.

\begin{table}[t]
    \centering
    \begin{tabular}{lccc}
        \toprule
        Model & Sample Rate & Model Size & MOS \\
        \midrule
       \textbf{\textit{VISinger 2}}  & 44.1KHz  & 25.7M & \(3.41 \pm 0.11\)        \\
       \textbf{\textit{SiFiSinger}}  & 44.1KHz & \textbf{22.5M} &\(\textbf{3.77} \pm \textbf{0.12}\)       \\
        \textbf{\textit{SiFiSinger-as}}  & 44.1KHz & 22.5M &\(3.64 \pm 0.13\)       \\
         \textbf{\textit{SiFiSinger-ds}}  & 44.1KHz & 22.5M &\(3.60 \pm 0.12\)       \\
        \midrule
        Recording   & 44.1KHz & - &\(4.23 \pm 0.07 \) \\
        \bottomrule
    \end{tabular}
    \caption{The MOS of \textbf{\textit{VISinger 2}} and \textbf{\textit{SiFiSinger}}-related systems.}
    \label{MOS table}
\end{table}
\vspace{-0.3cm}
\subsection{Subjective evaluation}
\vspace{-0.1cm}
{For the subjective evaluation, we evaluate the mean opinion score (MOS). To demonstrate the effectiveness of using the proposed modules in SiFiSinger, we  also perform a preference test on SiFiSinger and the two variants}. In each test, 30 audio segments synthesized by different systems  along with the ground truth (recordings) in the test set {are} randomly selected for evaluation by 10 experienced listeners.
%\subsubsection{MOS study}

{From Table \ref{MOS table},  we can see that \textbf{\textit{SiFiSinger}} and even the variants can significantly outperform \textbf{\textit{VISinger 2}} in terms of MOS. This {is caused by} the introduction of the source modules in both the prior encoder  and the decoder, the utilization of differentiable reconstruction loss as well as the decoupling relationship in the acoustic feature modeling process. 
 Furthermore, it is} worth noting that \textbf{\textit{SiFiSinger}} also has fewer model parameters compared to \textbf{\textit{VISinger 2}} due to the elimination of the DDSP synthesizer.
\begin{figure}[t]
    \centering
    \includegraphics[width=\columnwidth, keepaspectratio]{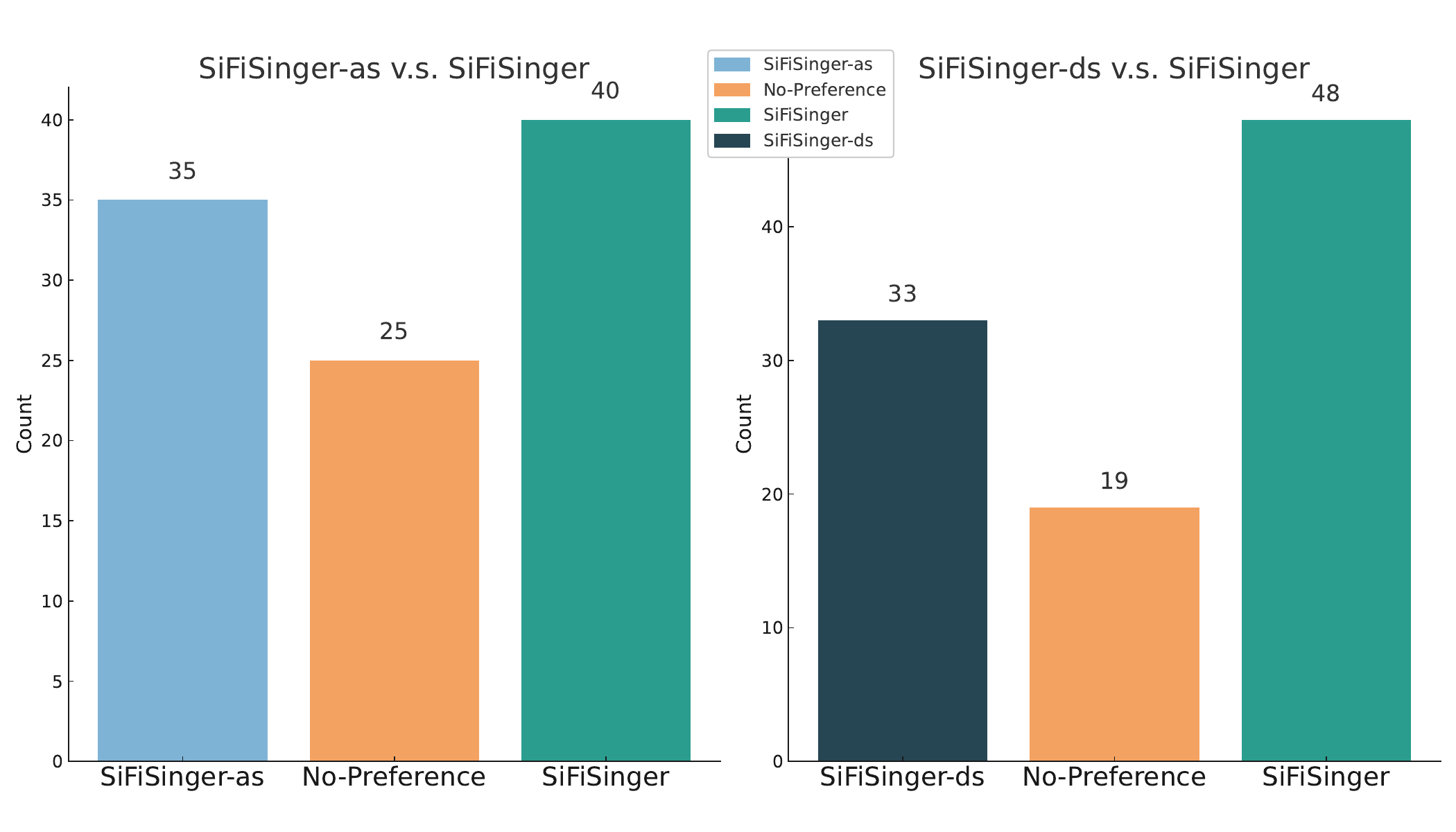}
    \caption{Preference test between \textbf{\textit{SiFiSinger}} and two ablation systems} 
    \label{fig3:ab_test}
    % \vspace{-0.3cm}
\end{figure}

{Finally, Fig~\ref{fig3:ab_test} shows an ablation study, where it is obvious that \textbf{\textit{SiFiSinger}} outperforms} the two ablation systems, showing the benefits of each component. Compar{ed to} \textbf{\textit{SiFiSinger-as}}, the AM source module in \textbf{\textit{SiFiSinger}} {can offer} improved harmonic components and pitch information, resulting in a better pitch accuracy and harmonic modeling. The differentiable reconstruction loss {can provide a} more direct supervision by back-propagating the gradient from the synthesized waveform, making {the synthesized audio quality of \textbf{\textit{SiFiSinger}} better than} that of \textbf{\textit{SiFiSinger-ds}}. 
 The difference between \textbf{\textit{SiFiSinger}} and  \textbf{\textit{SiFiSinger-ds}} is more significant than that between 
 \textbf{\textit{SiFiSinger}} and  \textbf{\textit{SiFiSinger-as}} as depicted in Fig.~\ref{fig3:ab_test}.
{Note that the MOS} value of \textbf{\textit{SiFiSinger-as}} is higher than \textbf{\textit{SiFiSinger-ds}}
{Altogether,  Table~\ref{MOS table} and Fig.~\ref{fig3:ab_test} reveals that the impact of removing the differentiable reconstruction loss is slightly greater than that of removing AM source m}odule.

\vspace{-0.3cm}
\section{CONCLUSION}
\label{sec:conclusion}
\vspace{-0.3cm}
In this paper, we {proposed SiFiSinger, an SVS system based neural source-filter,} which
eliminates the DDSP synthesizer in the VISinger 2 system and instead employs source modules to generate F0-controlled excitation signals. These signals are utilized at both the prior encoder and waveform decoder stages. The source modules integrate harmonic components using a learnable layer, allowing for more precise manipulation of pitch and tone. Additionally, we used differentiable methods to re-extract acoustic features from the synthesized waveform and computed the reconstruction loss {in order to provide a} more direct and effective supervision for the model. {The superiority of the proposed SiFiSinger  method lies in both pitch accuracy and audio quality compared to} VISinger 2 system.

\bibliographystyle{IEEEbib}
\bibliography{refs}

\end{document}